\documentclass[12pt]{article}

\def\square{\kern1pt\vbox{\hrule height 1.2pt\hbox{\vrule width 1.2pt\hskip 3pt
   \vbox{\vskip 6pt}\hskip 3pt\vrule width 0.6pt}\hrule height 0.6pt}\kern1pt}

\begin{document}

\begin{titlepage}

\begin{flushright}
UFIFT-QG-09-07
\end{flushright}

\vspace{1.5cm}

\begin{center}
{\bf The Volume of the Past Light-Cone and the Paneitz Operator}
\end{center}

\vspace{.5cm}

\begin{center}
Sohyun Park$^{\dagger}$ and R. P. Woodard$^{\ddagger}$
\end{center}

\vspace{.5cm}

\begin{center}
\it{Department of Physics \\
University of Florida \\
Gainesville, FL 32611}
\end{center}

\vspace{1cm}

\begin{center}
ABSTRACT
\end{center}
We study a conjecture involving the invariant volume of the past 
light-cone from an arbitrary observation point back to a fixed 
initial value surface. The conjecture is that a 4th order differential
operator which occurs in the theory of conformal anomalies gives $8\pi$
when acted upon the invariant volume of the past light-cone. We show
that an extended version of the conjecture is valid for an arbitrary 
homogeneous and isotropic geometry. First order perturbation theory 
about flat spacetime reveals a violation of the conjecture which, 
however, vanishes for any vacuum solution of the Einstein equation. 
These results may be significant for constructing quantum gravitational 
observables, for quantifying the back-reaction on spacetime expansion 
and for alternate gravity models which feature a timelike vector field.

\vspace{.5cm}

\begin{flushleft}
\end{flushleft}

\vspace{1.5cm}
\begin{flushleft}
$^{\dagger}$ e-mail: spark@phys.ufl.edu \\
$^{\ddagger}$ e-mail: woodard@phys.ufl.edu
\end{flushleft}
\end{titlepage}

\section{Introduction}

Suppose $\mathcal{S}$ is a Cauchy surface for the usual fields of
physics and let $\mathcal{M}$ stand for a globally hyperbolic 
spacetime manifold comprising $\mathcal{S}$ and its future. We 
will often think of $\mathcal{S}$ as the locus of points $x^{\mu} 
= (0,\vec{x})$, with $\mathcal{M}$ as the set of all $x^{\mu} = 
(t,\vec{x})$ with $t \geq 0$. Of course points are just labels, 
geometry derives from the metric field, $g_{\mu\nu}(t,\vec{x})$, 
which we shall take to be spacelike.

A quantity of great geometrical significance is the invariant volume 
of the past light-cone of an arbitrary point $x^{\mu} \in \mathcal{M}$.
It can be expressed as an integral involving some other geometrical
quantities which each require a little explanation,
\begin{equation}
\mathcal{V}[g](x) = \int_{\mathcal M} \!\! d^4x' \sqrt{-g(x')}
\, \Theta\Bigl(-\sigma[g](x,x')\Bigr) \Theta\Bigl(\mathcal{F}[g](x,x')
\Bigr) \; . \label{invV}
\end{equation}
(Our notation is that functional dependence upon fields appears
in square brackets, whereas dependence upon coordinates and other
parameters is parenthesized.) Of course $g(x')$ is the determinant
of $g_{\mu\nu}(x')$. The quantity $\sigma[g](x,x')$ was introduced 
by DeWitt and Brehme \cite{DWB}. It is one half the square of the
geodesic length from $x^{\mu}$ to $x^{\prime \mu}$ and can be 
expressed in terms of the geodesic $\chi^{\mu}[g](\tau,x,x')$ which 
runs between $x^{\mu}$ (at $\tau = 0$) to $x^{\prime \mu}$ (at 
$\tau = 1$),
\begin{equation}
\sigma[g](x,x') = \frac12 \int_0^1 \!\! d\tau \, g_{\mu\nu}(\chi)
\dot{\chi}^{\mu} \dot{\chi}^{\nu} \; . \label{sigma}
\end{equation}
If more than one geodesic connects $x^{\mu}$ and $x^{\prime \mu}$ 
then $\sigma[g](x,x')$ is defined to be the value for which the right
hand side of (\ref{sigma}) is smallest; if no geodesic connects the
two points then $\sigma[g](x,x')$ is $\frac12$ times the minimum 
distance between them. Because our metric is spacelike we see that 
$\sigma[g](x;x')$ is positive when $x^{\mu}$ and $x^{\prime \mu}$ 
are spacelike separated, and negative when they are timelike 
separated. The condition $\mathcal{F}[g](x,x') > 0$ in expression 
(\ref{invV}) restricts the integration to points $x^{\prime \mu}$ in 
the past of $x^{\mu}$. Owing to the factor of $\Theta(-\sigma)$ we 
need only define $\mathcal{F}[g](x,x')$ for the case where $x^{\mu}$ 
and $x^{\prime \mu}$ are timelike related: it is $+1$ when extending 
the geodesic to $\tau \geq 1$ eventually hits the Cauchy surface 
$\mathcal{S}$; otherwise it is $-1$.

The invariant volume of the past light-cone is interesting for a
number of reasons. First, if we consider $\mathcal{S}$ to be the
initial value surface (defined invariantly in some way) on which a 
quantum gravitational state is specified, then $\mathcal{V}[g](x)$ 
at some invariantly defined point $x^{\mu}$ ought to be an observable 
because a local observer should be able to look back into his past. It
is notoriously difficult to identify physically meaningful observables
in quantum gravity \cite{TW1,obs}. A second potential application is
quantifying the back-reaction to spacetime expansion. Suitable 
observables already exist for the important case of scalar-driven 
inflation \cite{phi} but these do not apply for pure quantum gravity 
and $\mathcal{V}[g](x)$ may have a role to play in invariantly fixing
the observation point \cite{TW2}. A final application concerns 
alternate gravity models which involve a timelike vector field 
\cite{Ted,JB}. Because $\mathcal{V}[g](x)$ necessarily grows as one 
evolves, its gradient is timelike, and can serve to define a timelike
vector field based upon the metric, without the complications
associated with introducing new dynamical degrees of freedom. It
has been suggested that such a term might arise from quantum corrections
to the effective field equations \cite{TW3}.

The purpose of this paper is to study a conjecture concerning 
$\mathcal{V}[g](x)$ and a certain 4th order differential operator. 
To motivate the conjecture, consider the flat space limit 
$g_{\mu\nu}(t,\vec{x}) \rightarrow \eta_{\mu\nu}$,
\begin{equation}
\sigma[\eta](x,x') = \frac12 (x \!-\! x')^2 \;\; , \;\;
\mathcal{F}[\eta](x,x') = {\rm sgn}(t \!-\! t') \;\; , \;\;
\mathcal{V}[\eta](x) = \frac{\pi}3 \, t^4 \; .
\end{equation}
Acting the square of the d'Alembertian ($\partial^2 \equiv \eta^{\mu\nu}
\partial_{\mu} \partial_{\nu}$) on $\mathcal{V}[\eta](x)$ gives
a simple constant,
\begin{equation}
\partial^4 \mathcal{V}[\eta](x) = 8 \pi \; . \label{flatconj}
\end{equation}
The conjecture is that a known differential operator $D_P$ allows us 
to extend relation (\ref{flatconj}) to an arbitrary, globally hyperbolic
metric (and for a general Cauchy surface $\mathcal{S}$),
\begin{equation}
D_P \mathcal{V}[g](x) = 8 \pi \; . \label{difconj}
\end{equation}
Of course there is no guarantee that any local differential operator
has this property. However, we will show that an extended version of
(\ref{difconj}) pertains for an arbitrary homogeneous and isotropic 
cosmology. We will also show that the conjecture fails for a general 
first order perturbation about flat spacetime, although only by terms 
which vanish with the vacuum Einstein equations. This suggests that 
some modified version of the conjecture might still be valid.

The Paneitz operator $D_P$ of our conjecture (\ref{difconj}) is known from 
the theory of conformal anomalies \cite{SD1,SD2}. For a general metric 
$g_{\mu\nu}(t,\vec{x})$ it takes the form,
\begin{equation}
D_P \equiv \square^2 + 2 D_{\mu} \Bigl[R^{\mu\nu} \!-\! \frac13 
g^{\mu\nu} R\Bigr] D_{\nu} \; , \label{Panop}
\end{equation}
where $R_{\mu\nu}$ is the Ricci tensor, $R$ is the Ricci scalar, 
$D_{\mu}$ is the covariant derivative operator and $\square$ is the 
covariant d`Alembertian,
\begin{equation}
\square \equiv g^{\mu\nu} D_{\mu} D_{\nu} \longrightarrow 
\frac1{\sqrt{-g}} \, \partial_{\mu} \Bigl[\sqrt{-g} g^{\mu\nu} 
\partial_{\nu} \Bigr] \qquad {\rm acting\ on\ a\ scalar.} 
\end{equation}
The Paneitz operator occurs in the nonlocal effective actions which
represent conformal anomalies \cite{SD1,SD2} owing to its special 
behavior under a conformal transformation,
\begin{equation}
g_{\mu\nu}(x) = \Omega^2(x) \widetilde{g}_{\mu\nu}(x) \qquad 
\Longrightarrow \qquad \Omega^{-4} \times \widetilde{D}_P \; .
\label{conformal}
\end{equation}
(Here $\widetilde{g}_{\mu\nu}$ is the conformally rescaled metric and 
$\widetilde{D}_P$ is the Paneitz operator constructed from it.) The
fact that all matter theories engender conformal anomalies means that
logarithms of $D_P$ are ubiquitous in the quantum effective action, and
inverses of $D_P$ must appear in the quantum-corrected, effective field 
equations. So our conjecture represents one way that the invariant
volume of the past light-cone can arise in the effective field 
equations of gravity without introducing new physics.

Just as Gauss's law has a differential and an integral form, so too
our conjecture (\ref{difconj}) can be expressed in terms of an integral.
The retarded Green's function $\mathcal{G}[g](x,x')$ of $D_P$ obeys,
\begin{equation}
\sqrt{-g} D_P \, \mathcal{G}[g](x;x') = \delta^4(x \!-\! x') \;\;
{\rm and} \;\; \Theta\Bigl(-\mathcal{F}[g](x,x')\Bigr) 
\mathcal{G}[g](x,x') = 0 \; . \label{P1}
\end{equation}
Because the characteristics of the highest derivative term in $D_P$
are set by the metric, in the same way as for typical second order 
operators, $\mathcal{G}[g](x,x')$ must vanish for any point $x^{\prime 
\mu}$ outside the past light-cone of $x^{\mu}$. Hence we should
get a finite result from integrating $\mathcal{G}[g](x,x')$ over 
$\mathcal{M}$ back to the initial value surface $\mathcal{S}$. We
define this integral as the functional $\mathcal{P}[g](x)$,
\begin{equation}
\mathcal{P}[g](x) \equiv \int_{\mathcal{M}} \!\! d^4x' \! \sqrt{-g(x')} \,
\mathcal{G}[g](x,x') \; , \label{P2}
\end{equation}
One can regard $\mathcal{P}[g](x)$ to be $D_P^{-1}$ acting on $1$,
so the integral expression of our conjecture (\ref{difconj}) is,
\begin{equation}
\mathcal{V}[g](x) = 8\pi \mathcal{P}[g](x) \; . \label{conj}
\end{equation}

In section 2 we demonstrate that an extended version of (\ref{conj}) 
pertains for an arbitrary homogeneous and isotropic geometry. In 
section 3 we consider the conjecture for first order perturbations 
about flat spacetime.  Although the conjecture is violated in general, 
it is valid for any first order perturbation which obeys the vacuum 
Einstein equations. We discuss the implications of this work in 
section 4. An appendix summarizes some useful but tedious integral 
identities.

\section{FRW Geometries}

The purpose of this section is to verify the conjecture (\ref{conj}) for
an arbitrary homogeneous and isotropic geometry,
\begin{equation}
\overline{g}_{\mu\nu}(t) dx^{\mu} dx^{\nu} = a^2(\eta) \Bigl[
-d\eta^2 + \frac{dr^2}{1 \!-\! k r^2} + r^2 d\Omega\Bigr] \equiv
a^2 \widetilde{g}_{\mu\nu} dx^{\mu} dx^{\nu} \; .
\end{equation}
Here $\eta$ is the conformal time, $a(\eta)$ is the scale factor and 
$k$ is the spatial curvature. The case of $k > 0$ corresponds to 
positive spatial curvature; $k=0$ is spatial flatness; and $k < 0$ is 
negative spatial curvature. For $k > 0$ it should be noted that $r$
has the finite range $0 \leq r \leq 1/\sqrt{k}$, and that any value 
of $r$ within this range corresponds to two distinct points on the
manifold. We first work out $\mathcal{V}[\overline{g}](x)$, then 
construct $\mathcal{P}[\overline{g}](x)$ and demonstrate that 
$\mathcal{V}[\overline{g}](x) = 8\pi \mathcal{P}[\overline{g}](x)$.

Because the geometry is homogeneous we can choose the spatial origin to
coincide with the point from which we are computing the invariant volume 
of the past light-cone. Recall that the invariant volume of the past 
light-cone from $x^{\mu} =(\eta,\vec{0})$ is the integral of $d^4x' 
\sqrt{-g(x')}$ over all points $x^{\prime \mu} = (\eta',\vec{x}')$ which 
are in the past of $x^{\mu}$ and timelike related to it. This obviously 
requires $\eta' < \eta$. To enforce the timelike relation we first 
compute the coordinate radius $r(\eta,\eta')$ which is traveled by a 
light ray emitted at $x^{\prime \mu}$ and received at $x^{\mu}$,
\begin{equation}
\int_0^r \!\! \frac{dx}{\sqrt{1 \!-\! k x^2}} = \int_{\eta'}^{\eta} \!\! ds
\qquad \Longrightarrow \qquad r(\eta,\eta') = \frac1{\sqrt{k}} \, 
\Bigl\vert \sin(\sqrt{k} \Delta \eta) \Bigr\vert \; . \label{radius}
\end{equation}
Here $\Delta \eta \equiv \eta - \eta'$, and we should call attention
to the fact that the formula for $r(\eta,\eta')$ remains valid no matter
what is the sign $k$. However, one should note that for $\sqrt{k} \, \Delta
\eta > \pi$ the light-cone has wrapped all the way around the spatial
manifold.

The points $x^{\prime \mu} = (\eta',\vec{x}')$ which are lightlike 
related to $x^{\mu} = (t,\vec{0})$ can be written as,
\begin{equation}
\vec{x}' = r(\eta,\eta') \times \widehat{r}(\theta',\phi') \; ,
\end{equation}
where the radial unit vector is the same as in flat space,
\begin{equation}
\widehat{r}(\theta'\phi') \equiv \Bigl( \sin(\theta') \cos(\phi'),
\sin(\theta') \sin(\phi'), \cos(\theta')\Bigr) \; .
\end{equation}
Suppose the initial value surface is at $\eta' = \eta_I$ and that, for the
positive curvature case, the observation time $\eta$ is not so late that
the light-cone has wrapped all the way around the spatial manifold. It 
follows that the invariant volume of the past light-cone (in the background 
geometry) is,
\begin{eqnarray}
\mathcal{V}[\overline{g}](t,\vec{0}) & = & \int_{\eta_I}^{\eta} \!\! d\eta' 
\!  \int \!\! d^3x' \, \sqrt{-\overline{g}(\eta',\vec{x}')} \, 
\Theta\Bigl(r(\eta,\eta') \!-\! r'\Bigr) \; , \\
& = & 4\pi \int_{\eta_I}^{\eta} \!\! d\eta' \, a^4(\eta') \times \int_0^{
r(\eta,\eta')} \!\! \frac{r^{\prime 2} dr'}{\sqrt{1 \!-\! k r^{\prime 2}}} 
\; , \\
& = & \frac{\pi}{k^{\frac32}} \int_{\eta_I}^{\eta} \!\! d\eta' a^4(\eta')
\Bigl[2 \sqrt{k} \Delta \eta \!-\! \sin(2 \sqrt{k} \Delta \eta)\Bigr] \;.
\label{Vbar}
\end{eqnarray}
For the case of positive curvature and $\sqrt{k} (\eta - \eta_I) > \pi$ 
the result is more complicated,
\begin{eqnarray}
\lefteqn{k > 0 \;\; {\rm and} \:\; \sqrt{k} (\eta \!-\! \eta_I) > \pi \qquad
\Longrightarrow } \nonumber \\
& & \mathcal{V} = \frac{\pi}{k^{\frac32}} \Biggl\{\int_{\eta -\pi/\sqrt{k}}^{
\eta} \!\!\!\!\!\!\!\!\!\!\! d\eta' a^4(\eta') \Bigl[2 \sqrt{k} \Delta 
\eta \!-\! \sin(2 \sqrt{k} \Delta \eta)\Bigr] + 2 \pi \int_{\eta_I}^{\eta - 
\pi/\sqrt{k}} \!\!\!\!\!\!\!\!\!\!\! d\eta' \, a^4(\eta') \Biggr\} . 
\qquad \label{Vbar'}
\end{eqnarray}
This falsifies the original conjecture, but a very simple extension of
it can be made for which (\ref{Vbar}) remains correct at all times. The
extension is just to redefine the ``the volume of the past light-cone'' to
mean the sum of the volumes of the past light-cone from the observation
point {\it and} from any focal points at which past-directed, null
geodesics from the observation point converge. Finally, note that expressions 
(\ref{Vbar}) and (\ref{Vbar'}) are valid as well for $\vec{x} \neq \vec{0}$ 
owing to the homogeneity of the geometry.

To construct the Paneitz operator on the background geometry we first
extract the conformal factor $\Omega = a(\eta)$ and exploit the simple
scaling rule (\ref{conformal}),
\begin{equation}
\overline{D}_P = \frac1{a^4} \, \widetilde{D}_P \; . \label{PP1}
\end{equation}
Recall that $\widetilde{D}_P$ is the Paneitz operator construced in 
the conformally related metric,
\begin{equation}
d\widetilde{s}^2 \equiv \widetilde{g}_{\mu\nu} dx^{\mu} dx^{\nu} =
-d\eta^2 + \frac{dr^2}{1 \!-\! k r^2} + r^2 d\Omega \; . \label{confFRW}
\end{equation}
Now consider the action of the scalar d'Alembertian on a function 
which depends only on the conformal time $\eta$,
\begin{equation}
\widetilde{\square} f(\eta) = \frac1{\sqrt{-\widetilde{g}}} \, 
\partial_{\mu} \Bigl(\sqrt{-\widetilde{g}} \, \widetilde{g}^{\mu\nu} 
\partial_{\nu} f\Bigr) = -\frac{d^2 f}{d\eta^2} \; . \label{PP2}
\end{equation}
To get the curvature part of the Paneitz operator we recall the
simple form of the Ricci tensor and its trace in the conformally
rescaled geometry (\ref{confFRW}),
\begin{equation}
\widetilde{R}^{00} = 0 \quad , \quad \widetilde{R}^{0j} = 0 \quad , 
\quad \widetilde{R}^{ij} = 2 k \widetilde{g}^{ij} \quad , \quad 
\widetilde{R} = 6 k \; .
\end{equation}
Now consider the action of the curvature terms on the same function
$f(\eta)$,
\begin{equation}
2 \widetilde{D}_{\mu} \Bigl[ \widetilde{R}^{\mu\nu} \!-\! \frac13 
\widetilde{g}^{\mu\nu} \widetilde{R}\Bigr] \widetilde{D}_{\nu} f(\eta)
= \frac2{\sqrt{-\widetilde{g}}} \, \partial_{\mu} \Bigl[ 
\sqrt{-\widetilde{g}} \Bigl(\widetilde{R}^{\mu\nu} \!-\! \frac13
\widetilde{g}^{\mu\nu} \widetilde{R}\Bigr) \partial_{\nu} f\Bigr]
= 4 k \frac{d^2 f}{d\eta^2} \; . \label{PP3}
\end{equation}
Combining relations (\ref{PP1}), (\ref{PP2}) and (\ref{PP3}) gives,
\begin{equation}
\overline{D}_P f(\eta) = \frac1{a^4} \Bigl(\frac{d}{d\eta}\Bigr)^2
\Bigl[\Bigl(\frac{d}{d\eta}\Bigr)^2 + 4 k\Bigr] f(\eta) \; .
\label{simple}
\end{equation}

Now recall from (\ref{P1}-\ref{P2}) that constructing $\mathcal{P}[g](x)$ 
amounts to solving the differential equation,
\begin{equation}
D_P \mathcal{P}[g](x) = 1 \; ,
\end{equation}
subject to retarded boundary conditions. From (\ref{simple}) we see that
this requires inverting the product of two second order, differential
operators. The associated retarded Green's functions are,
\begin{eqnarray}
\Bigl(\frac{d}{d \eta}\Bigr)^2 G_1(\eta,\eta') & \!\!\!= \!\!\!& 
\delta(\eta\!-\!\eta') \Longrightarrow G_1 = 
\theta(\eta \!-\! \eta') \!\times\! (\eta \!-\! \eta') \; , \\
\Bigl[\Bigl(\frac{d}{d \eta}\Bigr)^2 \!\!+\! 4 k\Bigr] G_2(\eta,\eta')
& \!\!\!=\!\!\! & \delta(\eta\!-\!\eta') \Longrightarrow G_2 = 
\theta(\eta \!-\! \eta') \!\times\! \frac{\sin[2 \sqrt{k} \, 
(\eta \!-\! \eta')]}{2 \sqrt{k}} \; . \qquad
\end{eqnarray}
It follows that the unique solution for $\mathcal{P}[\overline{g}](x)$ is,
\begin{eqnarray}
\mathcal{P}[\overline{g}](x) & = & \int_{\eta_I}^{\eta} \!\! d\eta' \,
G_2(\eta,\eta') \! \int_{\eta_I}^{\eta'} \!\! d\eta'' \, G_1(\eta',\eta'')
a^4(\eta'') \; , \\
& = & \int_{\eta_I}^{\eta} \!\! d\eta'' \, a^4(\eta'') \! 
\int_{\eta''}^{\eta} \!\! d\eta' \, G_1(\eta',\eta'') G_2(\eta,\eta') \; , \\
& = & \frac1{8 k^{\frac32}} \int_{\eta_I}^{\eta} \!\! d\eta' \,
a^4(\eta') \Bigl\{2 \sqrt{k} (\eta \!-\! \eta') \!-\! \sin[2\sqrt{k} \,
(\eta \!-\! \eta')] \Bigr\} \; . \qquad \label{Pbar}
\end{eqnarray}
Multiplying (\ref{Pbar}) by $8\pi$ gives precisely (\ref{Vbar}). Note that
expression (\ref{Pbar}) is correct for $\mathcal{P}[\overline{g}](x)$ for
all $k$ and $\eta$, whereas expression (\ref{Vbar}) must be replaced by
(\ref{Vbar'}) to give $\mathcal{V}[\overline{g}](x)$ for the case of positive
curvature and times so late that the light-cone has wrapped all the way
around the spatial manifold. Hence the conjecture will not remain valid
unless we modify the volume of the past light-cone to multiply count
points which have been multiply covered.

\section{Perturbations about Flat Spacetime}

The purpose of this section is to compare $\mathcal{V}[g](x)$ with
$8\pi \mathcal{P}[g](x)$ by using first order perturbation theory
around flat space. That means we write the metric as,
\begin{equation}
g_{\mu\nu}(t,\vec{x}) = \eta_{\mu\nu} + h_{\mu\nu}(t,\vec{x}) \; .
\end{equation}
The field $h_{\mu\nu}(t,\vec{x})$ is known as the {\it graviton field}.
By convention its indices are raised and lowered with the Lorentz metric,
\begin{equation}
h^{\mu}_{~\nu} \equiv \eta^{\mu\rho} h_{\rho\nu} \qquad , \qquad
h^{\mu\nu} \equiv \eta^{\mu\rho} \eta^{\nu\sigma} h_{\rho\sigma} \qquad
{\rm and} \qquad h \equiv \eta^{\mu\nu} h_{\mu\nu} \; .
\end{equation}
In the first subsection we work out $\mathcal{V}[\eta+h](x)$ at first order
in $h_{\mu\nu}$; the corresponding first order variation in $8\pi
\mathcal{P}[\eta+h](x)$ is derived in subsection~\ref{dP}. In the final 
subsection we reduce the difference of the two expressions to an invariant 
form.

\subsection{First order perturbation of $\mathcal{V}[\eta+h](x)$} \label{dV}

One computes the first order correction to $\mathcal{V}[g](x)$ from
expression (\ref{invV}) by expanding the measure factor and the theta 
function which enforces that $x^{\mu}$ and $x^{\prime \mu}$ are timelike 
separated,
\begin{eqnarray}
\sqrt{-g(x')} & \!=\! & 1 + \frac12 h(x') + O(h^2) \; , \\
\Theta\Bigl(-\sigma[g](x;x')\Bigr) & \!=\! & \Theta\Bigl( -\frac12 (x'\!-\!x)^2
\Bigr) - \delta\Bigl( \frac12 (x' \!-\! x)^2\Bigr) \, \delta \sigma(x;x')
+ O(h^2) \; . \qquad 
\end{eqnarray}
Note that there is no first order correction to the functional 
$\mathcal{F}[g](x;x')$ whose sign determines whether $x^{\prime \mu}$ is 
in the past ($\mathcal{F} = +1$) or future ($\mathcal{F} = -1$) of 
$x^{\mu}$. Indeed, it is not changed to any order in perturbation theory,
\begin{equation}
\mathcal{F}[g](x,x') = \rm{sgn}(t \!-\! t') \; .
\end{equation}

From expression (\ref{sigma}) we see that the variation of
$\sigma[g](x,x')$ about any metric consists of the metric variation,
plus the endpoint variation and a term proportional to the geodesic
equation,
\begin{eqnarray}
\lefteqn{\delta \sigma[g](x,x') = \frac12 \int_0^1 \!\! d\tau \, 
\delta g_{\mu\nu}(\chi) \dot{\chi}^{\mu} \dot{\chi}^{\nu} 
+ g_{\mu\nu}(\chi) \dot{\chi}^{\mu} \delta \chi^{\nu}
\Bigr\vert_0^1 } \nonumber \\
& & \hspace{5cm} - \int_0^1 \!\! d\tau \, g_{\mu\nu}(\chi) \Bigl[
\ddot{\chi}^{\mu} + \Gamma^{\mu}_{~ \rho\sigma}(\chi) \dot{\chi}^{\rho}
\dot{\chi}^{\sigma} \Bigr] \delta \chi^{\nu} \; . \qquad
\end{eqnarray}
The metric perturbation is just $\delta g_{\mu\nu} = h_{\mu\nu}$
and the other two terms vanish because the endpoints are fixed and
$\chi^{\mu}$ is a geodesic. The zeroth order geodesic is,
$\overline{\chi}^{\mu}(\tau) = x^{\mu} + (x' \!-\! x)^{\mu} \tau$, so
the first order correction to (\ref{invV}) is,
\begin{eqnarray}
\lefteqn{\delta \mathcal{V}(x) = \frac12 \int_{\mathcal{M}} \!\! d^4x' \, 
\Theta(t \!-\! t') \Theta\Bigl( -\frac12 (x' \!-\! x)^2\Bigr) \, h(x') } 
\nonumber \\
& & \hspace{-.5cm} \!\!-\frac12 \int_{\mathcal{M}} \!\!d^4x' \,\Theta(t\!-\!t') 
\delta\Bigl( \frac12 (x' \!-\! x)^2\Bigr) \!\! \int_0^1 \!\!\! d\tau \,
h_{\mu\nu}\Bigl(x \!+\! (x' \!-\!x)\tau\Bigr) (x' \!-\! x)^{\mu} (x' \!-\!
x)^{\nu} , \qquad \\
& & \hspace{-.7cm} = \frac12 \int_0^t \!\!\! dt' \int \!\! d^3x'
\, \Theta\Bigl( t \!-\! t' \!-\! \Vert \vec{x} \!-\! \vec{x}'\Vert\Bigr)
h(t',\vec{x}') -\frac12 \!\int_0^1 \!\!\! d\tau \!\! \int \!\! d^3x'
\frac{\Theta(t \!-\! \Vert \vec{x} \!-\! \vec{x}' \Vert)}{\Vert \vec{x} 
\!-\! \vec{x}'\Vert} \nonumber \\
& & \hspace{3.4cm} \times h_{\mu\nu}\Bigl(t \!-\! \Vert 
\vec{x} \!-\! \vec{x}'\Vert \tau, \vec{x} \!+\! (\vec{x}' \!-\! \vec{x}) 
\tau\Bigr) (x' \!-\! x)^{\mu} (x' \!-\! x)^{\nu} . \qquad \label{explicit}
\end{eqnarray}

Note that the temporal differences in (\ref{explicit}) contain no factors
of $\tau$,
\begin{equation}
(x' \!-\! x)^0 \equiv -\Vert \vec{x}' \!-\! \vec{x}\Vert \equiv -\Delta x \; .
\end{equation}
So expanding out the double contraction in (\ref{explicit}) gives,
\begin{equation}
h_{\mu\nu}\Bigl(t \!-\! \Delta x \tau,\vec{x} \!+\! \Delta x \tau \widehat{r} 
\Bigr) (x' \!-\! x)^{\mu} (x' \!-\! x)^{\nu} = \Delta x^2 \Bigl\{ h_{00} - 
2 h_{0i} \widehat{r}^i + h_{ij} \widehat{r}^i \widehat{r}^j \Bigr\} \; .
\end{equation}
Here and subsequently the radial unit vector is,
\begin{equation}
\widehat{r} \equiv \frac{\vec{x}' \!-\! \vec{x}}{\Delta x} \; .
\end{equation}
The final form is obtained by changing variables in the second term of
(\ref{explicit}) from $\tau$ to the retarded time,
\begin{equation}
\tau \equiv \frac{t \!-\! t'}{\Delta x} \equiv \frac{\Delta t}{\Delta x}
\qquad \Longleftrightarrow \qquad t' \equiv t - \Delta x \, \tau \; .
\end{equation}
This allows us to perform the radial integration,
\begin{eqnarray}
\lefteqn{\int_0^1 \!\! d\tau \!\! \int \!\! d^3x' \, \Theta(t \!-\! \Delta x) 
\Delta x f\Bigl(t \!-\! r \tau,\vec{x} \!+\! \Delta x \tau \widehat{r}\Bigr) }
\nonumber \\
& & \hspace{5cm} = \int \!\! d\Omega \!\! \int_0^t \!\! dr \, r^3 \!\! 
\int_0^1 \!\! d\tau \, f\Bigl(t \!-\! r \tau,\vec{x} \!+\! \Delta x \tau 
\widehat{r}\Bigr) \; , \qquad \\
& & \hspace{5cm} = \int \!\! d\Omega \!\! \int_0^t \!\! dr \, r^2 \!\! \int_{t-
r}^t \!\!  dt' \, f\Bigl(t',\vec{x} \!+\! \Delta t \, \widehat{r}\Bigr) \; , \\
& & \hspace{5cm} = \int_0^t \!\! dt' \!\! \int \!\! d\Omega \, 
f\Bigl(t',\vec{x} \!+\! \Delta t \, \widehat{r}\Bigr) \!\! \int_{\Delta t}^t 
\!\! dr \, r^2 \; , \\
& & \hspace{5cm} = \frac13 \int_0^t \!\! dt' \, (t^3 \!-\! \Delta t^3) \!\! 
\int \!\! d\Omega \, f\Bigl(t',\vec{x} \!+\! \Delta t \, \widehat{r}\Bigr) \; .
\end{eqnarray}
Hence our final form for the first order perturbation of 
$\mathcal{V}[g](x)$ is,
\begin{eqnarray}
\lefteqn{\delta \mathcal{V}(x) = \frac12 \int_0^t \!\!\! dt' \int \!\! d^3x'
\, \Theta\Bigl( \Delta t \!-\! \Delta x\Bigr) h(t',\vec{x}') -\frac16
\! \int_0^t \!\! dt' (t^3 \!-\! \Delta t^3) \!\! \int \!\! d\Omega }
\nonumber \\
& & \hspace{.7cm} \times \Biggl\{ 
h_{00}\Bigl(t',\vec{x} \!+\! \Delta t \, \widehat{r}\Bigr) 
- 2 h_{0i}\Bigl(t',\vec{x} \!+\! \Delta t \, \widehat{r}\Bigr) \widehat{r}^i
+ h_{ij}\Bigl(t',\vec{x} \!+\! \Delta t \, \widehat{r}\Bigr) \widehat{r}^i
\widehat{r}^j \Biggr\} . \qquad \label{Volfin}
\end{eqnarray}

\subsection{First order perturbation of $8\pi \mathcal{P}[\eta+h](x)$} 
\label{dP}

Recall that $\mathcal{P}[g](x)$ can be expressed as the inverse of the 
Paneitz operator acting on unity,
\begin{equation}
\mathcal{P}[g](x) \equiv \int_{\mathcal{M}} \!\! d^4x' \sqrt{-g(x')} \, 
\mathcal{G}[g](x,x') = \frac1{D_P} \Bigl[1\Bigr](x) \; ,
\end{equation}
If we write,
\begin{equation}
D_P = \overline{D_P} + \delta D_P + O(h^2) \; ,
\end{equation}
then the functional inverse becomes,
\begin{equation}
\frac1{D_P} = \frac1{\overline{D_P}} - \frac1{\overline{D_P}} \times 
\delta D_P \times \frac1{\overline{D_P}} + O(h^2) \; . \label{DPexp}
\end{equation}
The first order correction we are seeking is accordingly,
\begin{eqnarray}
\delta \mathcal{P}(x) & = & -\int_{\mathcal{M}} \!\! d^4x' \, 
\mathcal{G}[\eta](x,x') \times \delta D_P' \times \mathcal{P}[\eta](x') \; , \\
& = & -\int_{\mathcal{M}} \!\! d^4x' \, \frac1{8\pi} \, \Theta(t \!-\! t')
\Theta\Bigl(-(x \!-\! x')^2\Bigr) \times \delta D_P' \times \frac1{24} \,
t^{\prime 4} \; . \label{deltaP}
\end{eqnarray}

It remains to work out the first order variation of the Paneitz operator
(\ref{Panop}). Because the Ricci tensor vanishes for flat space the 
background value of the Paneitz operator is just the square of the flat 
space d'Alembertian,
\begin{equation}
\overline{D_P} = \Bigl(\partial^2\Bigr)^2 \; .
\end{equation}
Expanding the scalar d'Alembertian in powers of the graviton field gives,
\begin{equation}
\square \equiv \frac1{\sqrt{-g}} \, \partial_{\mu} \Bigl[\sqrt{-g} 
g^{\mu\nu} \partial_{\nu} \Bigr] = \partial^2 + \frac12 h^{,\mu} 
\partial_{\mu} - \partial_{\mu} h^{\mu\nu} \partial_{\nu} + O(h^2) \; .
\end{equation}
Therefore the expansion of $\square^2$ is,
\begin{equation}
\square^2 = \partial^4 + \partial^2 \Bigl[ \frac12 h^{,\mu} \partial_{\mu}
\!-\! \partial_{\mu} h^{\mu\nu} \partial_{\nu}\Bigr] + \Bigl[ \frac12 
h^{,\mu} \partial_{\mu} \!-\! \partial_{\mu} h^{\mu\nu} \partial_{\nu}\Bigr]
\partial^2 + O(h^2) \; . \label{term1}
\end{equation}

The Riemann tensor is first order in the graviton field,
\begin{equation}
R_{\rho\sigma\mu\nu} = -\frac12 \Bigl( h_{\rho\mu , \sigma\nu} - 
h_{\mu\sigma , \nu\rho} + h_{\sigma\nu , \rho \mu} - h_{\nu\rho , \mu\sigma}
\Bigr) + O(h^2) \; .
\end{equation}
Hence the expansions of the Ricci tensor and the Ricci scalar are,
\begin{eqnarray}
R_{\mu\nu} & = & \frac12 \Bigl( h^{\rho}_{~\mu , \nu \rho} + 
h^{\rho}_{~\nu , \mu \rho} - h_{, \mu\nu} - h_{\mu\nu , ~\rho}^{~~~\rho}\Bigr) 
+ O(h^2) \; , \\
R & = & h^{\rho\sigma}_{~~ ,\rho\sigma} - h^{,\rho}_{~~\rho} + O(h^2) \; .
\end{eqnarray}
Because the curvature terms are already first order in the graviton field we
do not need to worry about the distinction between covariant differentiation
and ordinary differentiation in computing the expansions of the two
curvature terms in the Paneitz operator,
\begin{eqnarray}
2 D_{\mu} R^{\mu\nu} D_{\nu} & = & \partial_{\mu} \Bigl(h^{\rho \mu , \nu\rho} 
\!+\! h^{\rho\nu , \mu\rho} \!-\! h^{,\mu\nu} \!-\! h^{\mu\nu , 
\rho}_{~~~~ \rho} \Bigr) \partial_{\nu} + O(h^2) \; , \label{term2} \\
-\frac23 D_{\mu} g^{\mu} R D_{\nu} & = & -\frac23 \partial_{\mu} \Bigl(
h^{\rho\sigma}_{~~ ,\rho\sigma} \!-\! h^{, \rho}_{~~ \rho} \Bigr) 
\partial^{\mu} + O(h^2) \; . \label{term3}
\end{eqnarray}

\begin{table}

\vbox{\tabskip=0pt \offinterlineskip
\def\tablerule{\noalign{\hrule}}
\halign to390pt {\strut#& \vrule#\tabskip=1em plus2em&
\hfil#\hfil& \vrule#& \hfil#\hfil& \vrule#& \hfil#\hfil& 
\vrule#& \hfil#\hfil& \vrule#\tabskip=0pt\cr \tablerule
\omit&height4pt&\omit&&\omit&&\omit&&\omit&\cr
\omit&height2pt&\omit&&\omit&&\omit&&\omit&\cr 
&&\omit\hidewidth ${\rm I}$ \hidewidth 
&& $\!\!\!\! (\delta D_P)_I \!\!\!\!$ &&
\hidewidth ${\rm I}$ \hidewidth 
&& $\!\!\!\! (\delta D_P)_I \!\!\!\!$ & \cr
\omit&height4pt&\omit&&\omit&&\omit&&\omit&\cr \tablerule 
\omit&height2pt&\omit&&\omit&&\omit&&\omit&\cr 
&& 1 && $+\frac12 \partial^2 h^{,\mu} \partial_{\mu}$ 
&& 6 && $+\partial_{\mu} h^{\rho \nu , \mu}_{~~~~ \rho} \partial_{\nu}$ & \cr
\omit&height2pt&\omit&&\omit&&\omit&&\omit&\cr \tablerule
\omit&height2pt&\omit&&\omit&&\omit&&\omit&\cr 
&& 2 && $-\partial^2 \partial_{\mu} h^{\mu\nu} \partial_{\nu}$
&& 7 && $-\partial_{\mu} h^{,\mu\nu} \partial_{\nu}$ & \cr
\omit&height2pt&\omit&&\omit&&\omit&&\omit&\cr \tablerule
\omit&height2pt&\omit&&\omit&&\omit&&\omit&\cr 
&& 3 && $+\frac12 h^{,\mu} \partial_{\mu} \partial^2$ 
&& 8 && $-\partial_{\mu} h^{\mu \nu , \rho}_{~~~~ \rho} \partial_{\nu}$ & \cr
\omit&height2pt&\omit&&\omit&&\omit&&\omit&\cr \tablerule
\omit&height2pt&\omit&&\omit&&\omit&&\omit&\cr 
&& 4 && $- \partial_{\mu} h^{\mu\nu} \partial_{\nu} \partial^2$ 
&& 9 && $- \frac23 \partial_{\mu} h^{\rho \sigma}_{~~ , \rho\sigma} 
\partial^{\mu}$ & \cr
\omit&height2pt&\omit&&\omit&&\omit&&\omit&\cr \tablerule
\omit&height2pt&\omit&&\omit&&\omit&&\omit&\cr 
&& 5 && $+ \partial_{\mu} h^{\rho \mu , \nu}_{~~~~ \rho} \partial_{\nu}$ 
&& 10 && $+\frac23 \partial_{\mu} h^{, \rho}_{~~ \rho} \partial^{\mu}$ & \cr
\omit&height2pt&\omit&&\omit&&\omit&&\omit&\cr \tablerule}}

\caption{First order perturbations of the Paneitz operator.}

\label{dD_P}

\end{table}

Adding the first order contributions from expressions (\ref{term1})
and (\ref{term2}-\ref{term3}) gives $\delta D_P$,
\begin{eqnarray}
\lefteqn{\delta D_P = \partial^2 \Bigl[ \frac12 h^{,\mu} \partial_{\mu}
\!-\! \partial_{\mu} h^{\mu\nu} \partial_{\nu}\Bigr] + \Bigl[ \frac12 
h^{,\mu} \partial_{\mu} \!-\! \partial_{\mu} h^{\mu\nu} \partial_{\nu}\Bigr]
\partial^2 } \nonumber \\
& & \hspace{1.5cm} + \partial_{\mu} \Bigl(h^{\rho \mu , \nu}_{~~~~\rho} 
\!+\!  h^{\rho\nu , \mu}_{~~~~\rho} \!-\! h^{,\mu\nu} \!-\! 
h^{\mu\nu , \rho}_{~~~~ \rho} \Bigr) \partial_{\nu} -\frac23 
\partial_{\mu} \Bigl(h^{\rho\sigma}_{~~ ,\rho \sigma} \!-\! 
h^{, \rho}_{~~ \rho} \Bigr) \partial^{\mu} \; . \qquad \label{deltaD}
\end{eqnarray}
We have assigned each of the ten operators of (\ref{deltaD}) an arbitrary 
number and listed them in Table~\ref{dD_P}. We shall employ this notation,
$(\delta D)_I$ for $I$ from 1 to 10, in the reductions of the subsequent
subsection.

\subsection{The deficit term} \label{deficit}

Recall that expression (\ref{Volfin}) for $\delta \mathcal{V}(x)$ gives 
the first order perturbation of the left hand side of our conjecture 
(\ref{conj}). Combining equations (\ref{deltaP}) and (\ref{deltaD}) from 
the previous subsection gives an expression for the first order 
perturbation of the right hand side,
\begin{equation}
8\pi \delta \mathcal{P}(x) = -\frac1{24} \int_0^t \!\! dt' \!\! \int \!\! d^3x'
\, \Theta\Bigl( \Delta t \!-\! \Delta x\Bigr) \, \sum_{I=1}^{10}
\Bigl(\delta D_P' \Bigr)_I \, t^{\prime 4} \; , \label{Panfin}
\end{equation}
where $\Delta t \equiv t \!-\! t'$, $\Delta x \equiv \Vert \vec{x} \!-\!
\vec{x}'\Vert$, and the operators $(\delta D_P)_I$ are listed in 
Table~\ref{dD_P}. Although (\ref{Volfin}) and (\ref{Panfin}) are correct
and complete, it is not obvious whether or not they agree. To compare them
we will reduce (\ref{Panfin}) to the same form as (\ref{Volfin}). This
can be accomplished by the following steps:
\begin{enumerate}
\item{Act any derivatives from $(\delta D_P')_I$ which stand to the right
of the $h_{\mu\nu}(x')$ on the factor of $t^{\prime 4}$; then}
\item{Integrate by parts to remove all the derivatives from the graviton
fields.}
\end{enumerate}
Step 2 produces volume terms which are integrated throughout the 
light-cone and surface terms restricted to its boundary. If (\ref{conj}) 
is correct then the sum of all the volume terms must agree with the first 
integral of (\ref{Volfin}), and the sum of all the surface terms must 
agree with the second integral of (\ref{Volfin}).

\begin{table}

\vbox{\tabskip=0pt \offinterlineskip
\def\tablerule{\noalign{\hrule}}
\halign to390pt {\strut#& \vrule#\tabskip=1em plus2em&
\hfil#\hfil& \vrule#& \hfil#\hfil& \vrule#& \hfil#\hfil& \vrule#& \hfil#\hfil&
\vrule#\tabskip=0pt\cr \tablerule
\omit&height4pt&\omit&&\omit&&\omit&&\omit&\cr
\omit&height2pt&\omit&&\omit&&\omit&&\omit&\cr 
&&\omit\hidewidth $\!\!\!\! \; {\rm \#} \!\!\!\!$ \hidewidth 
&& $\!\!\!\!\! {\rm Coef.\ of} \; h_{00} \!\!\!\!\!$ 
&& $ {\rm Coef.\ of} \; \widehat{r}^i h_{00,i} $ 
&& $\!\! {\rm Coef.\ of} \; \widehat{r}^i \widehat{r}^j h_{00,ij} \!\!$ 
& \cr \omit&height4pt&\omit&&\omit&&\omit&&\omit&\cr \tablerule 
\omit&height2pt&\omit&&\omit&&\omit&&\omit&\cr 
&& $\!\!\!\! 1 \!\!\!\!$ 
&& $\!\!\!\!\! \frac16 t^{\prime 3} - \frac12 t^{\prime 2} \Delta t \!\!\!\!\!$ 
&& $\frac16 t^{\prime 3} \Delta t$ 
&& $\!\! 0 \!\!$ & \cr
\omit&height2pt&\omit&&\omit&&\omit&&\omit&\cr \tablerule
\omit&height2pt&\omit&&\omit&&\omit&&\omit&\cr 
&& $\!\!\!\! 2 \!\!\!\!$ 
&& $\!\!\!\!\! -\frac13 t^{\prime 3}$ && $-\frac13 t^{\prime 3} \Delta t \!\!\!\!\!$ 
&& $\!\! 0 \!\!$ 
& \cr \omit&height2pt&\omit&&\omit&&\omit&&\omit&\cr \tablerule
\omit&height2pt&\omit&&\omit&&\omit&&\omit&\cr 
&& $\!\!\!\! 3 \!\!\!\!$ 
&& $\!\!\!\!\! \frac12 t^{\prime } \Delta t^2 \!\!\!\!\!$ 
&& $0$ 
&& $\!\! 0 \!\!$ 
& \cr
\omit&height2pt&\omit&&\omit&&\omit&&\omit&\cr \tablerule
\omit&height2pt&\omit&&\omit&&\omit&&\omit&\cr 
&& $\!\!\!\! 4 \!\!\!\!$ 
&& $\!\!\!\!\! -t^{\prime } \Delta t^2 \!\!\!\!\!$ 
&& $0$ 
&& $\!\! 0 \!\!$ 
& \cr \omit&height2pt&\omit&&\omit&&\omit&&\omit&\cr \tablerule
\omit&height2pt&\omit&&\omit&&\omit&&\omit&\cr 
&& $\!\!\!\! 5 \!\!\!\!$ 
&& $\!\!\!\!\! \frac13 t^{\prime 3} - 2 t^{\prime 2} \Delta t +t^{\prime } \Delta t^2 \!\!\!\!\!$ 
&& $\frac23 t^{\prime 3} \Delta t -t^{\prime 2} \Delta t^2$ 
&& $\!\! \frac1{6} t^{\prime 3} \Delta t^2 \!\!$ & 
\cr \omit&height2pt&\omit&&\omit&&\omit&&\omit&\cr \tablerule
\omit&height2pt&\omit&&\omit&&\omit&&\omit&\cr 
&& $\!\!\!\! 6 \!\!\!\!$ 
&& $\!\!\!\!\! \frac13 t^{\prime 3} - 2 t^{\prime 2} \Delta t +t^{\prime } \Delta t^2 \!\!\!\!\!$ 
&& $\frac13 t^{\prime 3} \Delta t -\frac12 t^{\prime 2} \Delta t^2$ 
&& $\!\! 0 \!\!$ & \cr
\omit&height2pt&\omit&&\omit&&\omit&&\omit&\cr \tablerule
\omit&height2pt&\omit&&\omit&&\omit&&\omit&\cr 
&& $\!\!\!\! 7 \!\!\!\!$ 
&& $\!\!\!\!\!-\frac13 t^{\prime 3} + 2 t^{\prime 2} \Delta t - t^{\prime } \Delta t^2\!\!\!\!\!$ 
&& $-\frac13 t^{\prime 3} \Delta t +\frac12 t^{\prime 2} \Delta t^2$ 
&& $\!\! 0 \!\!$ 
& \cr \omit&height2pt&\omit&&\omit&&\omit&&\omit&\cr \tablerule
\omit&height2pt&\omit&&\omit&&\omit&&\omit&\cr 
&& $\!\!\!\! 8 \!\!\!\!$ 
&& $\!\!\!\!\!-\frac13 t^{\prime 3} + 2 t^{\prime 2} \Delta t - t^{\prime } \Delta t^2\!\!\!\!\!$ 
&& $-\frac13 t^{\prime 3} \Delta t + t^{\prime 2} \Delta t^2$ 
&& $\!\! 0 \!\!$ 
& \cr \omit&height2pt&\omit&&\omit&&\omit&&\omit&\cr \tablerule
\omit&height2pt&\omit&&\omit&&\omit&&\omit&\cr 
&& $\!\!\!\! 9 \!\!\!\!$ 
&& $\!\!\!\!\! -\frac29 t^{\prime 3} + \frac43 t^{\prime 2} \Delta t -\frac23 t^{\prime } \Delta t^2\!\!\!\!\!$ 
&& $-\frac49 t^{\prime 3} \Delta t +\frac23 t^{\prime 2} \Delta t^2$ 
&& $\!\! -\frac1{9} t^{\prime 3} \Delta t^2 \!\!$ & 
\cr \omit&height2pt&\omit&&\omit&&\omit&&\omit&\cr \tablerule
\omit&height2pt&\omit&&\omit&&\omit&&\omit&\cr 
&& $\!\!\!\! 10 \!\!\!\!$ 
&& $\!\!\!\!\! +\frac29 t^{\prime 3} - \frac43 t^{\prime 2} \Delta t +\frac23 t^{\prime } \Delta t^2\!\!\!\!\!$ 
&& $\frac29 t^{\prime 3} \Delta t -\frac23 t^{\prime 2} \Delta t^2$ 
&& $\!\! 0 \!\!$ 
& \cr
\omit&height2pt&\omit&&\omit&&\omit&&\omit&\cr \tablerule
\omit&height1pt&\omit&&\omit&&\omit&&\omit&\cr \tablerule
\omit&height2pt&\omit&&\omit&&\omit&&\omit&\cr 
&& $\!\!\!\!{\rm Sum}\!\!\!\!\!\!$ 
&& $\!\!\!\!\! -\frac16 t^{\prime 3} -\frac12 t^{\prime 2} \Delta t -\frac12 t' \Delta t^2 \!\!\!\!\!$ 
&& $-\frac1{18} t^{\prime 3} \Delta t$ 
&& $\!\! \frac1{18} t^{\prime 3} \Delta t^2 \!\!$ & 
\cr \omit&height2pt&\omit&&\omit&&\omit&&\omit&\cr \tablerule}}

\caption{Reductions involving $h_{00}$. Each coefficient appears
in the form $\int_0^t \! dt' \! \int \! d\Omega \times {\rm Coef.} \times
f(t',\vec{x} \!+\! \Delta t \, \widehat{r})$.}

\label{h00}

\end{table}

It turns out that only $(D_P)_3$ produces a volume term, and
this volume term agrees with the first integral in (\ref{Volfin}).
Tables~\ref{h00}-\ref{hijij} summarize our results for the surface
terms. To illustrate the reduction procedure consider $(D_P)_1 =
\frac12 \partial^2 h^{,\mu} \partial_{\mu}$. Step 1 gives,
\begin{eqnarray}
\lefteqn{-\frac1{24} \! \int_0^t \!\! dt' \!\! \int \!\! d^3x' \, 
\Theta(\Delta t \!-\! \Delta x) \Bigl[-\frac12 \partial^{\prime 2} 
\dot{h}(t',\vec{x}') \partial_0'\Bigr] t^{\prime 4} } \nonumber \\
& & \hspace{4.5cm} = \frac1{12} \! \int_0^t \!\! dt' \!\! \int 
\!\! d^3x' \, \Theta(\Delta t \!-\! \Delta x) \partial^{\prime 2} 
\Bigl[\dot{h}(t',\vec{x}') t^{\prime 3} \Bigr] \; . \qquad \label{step1}
\end{eqnarray}

The next step is to partially integrate the $\partial^{\prime 2}$.
It would be silly to act this on the $\dot{h}(t',\vec{x}') t^{\prime 3}$
because we must throw all derivatives off the graviton field in order 
to reach the same form as (\ref{Volfin}). So we instead partially 
integrate it immediately. Note also that the only surface terms lie
on the boundary of the light-cone:
\begin{itemize}
\item{Surface terms at spatial infinity are zero from the
$\Theta(\Delta t \!-\! \Delta x)$;}
\item{Surface terms at $t' = 0$ vanish on account of the factor of 
$t^{\prime 3}$; and}
\item{Surface terms at $t' = t$ vanish because the theta function 
becomes $\Theta(0 \!-\! \Delta x)$, which restricts $\vec{x}'$ to a 
region of zero volume around $\vec{x}$.}
\end{itemize}
The only contribution comes from when the $\partial^{\prime 2}$ 
acts on the theta function,
\begin{equation}
\partial^{\prime 2} \Theta(\Delta t \!-\! \Delta x) = -\frac2{\Delta x} \, 
\delta(\Delta t \!-\! \Delta x) \; . \label{result}
\end{equation}

Substituting (\ref{result}) in (\ref{step1}) gives,
\begin{eqnarray}
\lefteqn{-\frac1{24} \! \int_0^t \!\! dt' \!\! \int \!\! d^3x' \, 
\Theta(\Delta t \!-\! \Delta x) \Bigl[-\frac12 \partial^{\prime 2} 
\dot{h}(t',\vec{x}') \partial_0'\Bigr] t^{\prime 4} } \nonumber \\
& & \hspace{3.5cm} = -\frac16 \! \int_0^t \!\! dt' \, t^{\prime 3} \!\! 
\int \!\! d\Omega \!\! \int_0^{\infty} \!\!\! dr \, r 
\delta(\Delta t \!-\! r) \dot{h}\Bigl(t',\vec{x} \!+\! r \widehat{r}\Bigr) 
\; , \qquad \\
& & \hspace{3.5cm} = -\frac16 \! \int_0^t \!\! dt' \, t^{\prime 3} \Delta t
\! \int \!\! d\Omega \, \dot{h}\Bigl(t',\vec{x} \!+\! \Delta t \, 
\widehat{r} \Bigr) \; . \qquad \label{step2}
\end{eqnarray}
Note that the time derivative in $\dot{h}(t',\vec{x} \!+\! \Delta t \, 
\widehat{r})$ in expression (\ref{step2}) is only with respect to the first 
argument; it does not include the $t'$ dependence of $\Delta t = t \!-\! t'$
in the spatial argument. The full derivative with respect to $t'$ is,
\begin{equation}
\frac{\partial}{\partial t'} \, h\Bigl(t',\vec{x} \!+\! \Delta t \, 
\widehat{r}\Bigr) = \dot{h}\Bigl(t',\vec{x} \!+\! \Delta t \, 
\widehat{r}\Bigr) - \widehat{r} \!\cdot\! \vec{\nabla} \, h\Bigl(t',\vec{x} 
\!+\! \Delta t \, \widehat{r}\Bigr) . \label{fulld}
\end{equation}
The final result is,
\begin{eqnarray}
\lefteqn{-\frac1{24} \! \int_0^t \!\! dt' \!\! \int \!\! d^3x' \, 
\Theta(\Delta t \!-\! \Delta x) \Bigl[-\frac12 \partial^{\prime 2} 
\dot{h}(t',\vec{x}') \partial_0'\Bigr] t^{\prime 4} 
= \int_0^t \!\! dt' \Bigl[-\frac16 t^{\prime 3} \!+\! \frac12 t^{\prime 2} 
\Delta t\Bigr] } \nonumber \\
& & \hspace{1.5cm} \times \! \int \!\! d\Omega \, 
h\Bigl(t',\vec{x} \!+\! \Delta t \, \widehat{r} \Bigr) 
- \frac16 \!\! \int_0^t \!\! dt' \, t^{\prime 3} \Delta t \!\! \int \!\! 
d\Omega \, \widehat{r} \!\cdot\! \vec{\nabla} \, h\Bigl(t',\vec{x} \!+\! 
\Delta t \, \widehat{r} \Bigr) \; . \qquad \label{finalstep}
\end{eqnarray}
Upon substituting the $3+1$ decomposition $h = -h_{00} + h_{ii}$ we
have the first row of entries for Tables~\ref{h00} and \ref{hii}.

\begin{table}

\vbox{\tabskip=0pt \offinterlineskip
\def\tablerule{\noalign{\hrule}}
\halign to390pt {\strut#& \vrule#\tabskip=1em plus2em&
\hfil#\hfil& \vrule#& \hfil#\hfil& \vrule#& \hfil#\hfil&
\vrule#& \hfil#\hfil& \vrule#\tabskip=0pt\cr \tablerule
\omit&height4pt&\omit&&\omit&&\omit&&\omit&\cr
\omit&height2pt&\omit&&\omit&&\omit&&\omit&\cr 
&&\omit\hidewidth $\!\!\!\!\!\!\;{\rm \#}\!\!\!\!\!\!$ \hidewidth 
&& $\!\!\!\!\!\!{\rm Coef.\ of} \; \widehat{r}^i h_{0i} \!\!\!\!\!\!\!$ 
&& $\!\!\!\!\!\!{\rm Coef.\ of} \; h_{0i,i} \!\!\!\!\!\!\!$ 
&& $\!\!\!\!\!{\rm Coef.\ of} \; \widehat{r}^i h_{0j,ij} \!\!\!\!\!$ & \cr 
\omit&height4pt&\omit&&\omit&&\omit&&\omit&\cr \tablerule 
\omit&height2pt&\omit&&\omit&&\omit&&\omit&\cr 
&& $\!\!\!\!\!\!1\!\!\!\!\!\!$ && $\!\!\!\!\!\!0\!\!\!\!\!\!$ 
&& $0$ && $0$ & \cr
\omit&height2pt&\omit&&\omit&&\omit&&\omit&\cr \tablerule
\omit&height2pt&\omit&&\omit&&\omit&&\omit&\cr 
&& $\!\!\!\!\!\!2\!\!\!\!\!\!$ && $\!\!\!\!\!\!0\!\!\!\!\!\!$ 
&& $\!\!\!\!\!\! \frac13 t^{\prime 3} \Delta t \!\!\!\!\!\! $ && $0$ & \cr
\omit&height2pt&\omit&&\omit&&\omit&&\omit&\cr \tablerule
\omit&height2pt&\omit&&\omit&&\omit&&\omit&\cr 
&& $\!\!\!\!\!\!3\!\!\!\!\!\!$ && $\!\!\!\!\!\!0\!\!\!\!\!\!$ 
&& $\!\!\!\!\!\!0\!\!\!\!\!\!$ && $0$ & \cr
\omit&height2pt&\omit&&\omit&&\omit&&\omit&\cr \tablerule
\omit&height2pt&\omit&&\omit&&\omit&&\omit&\cr 
&& $\!\!\!\!\!\!4\!\!\!\!\!\!$ 
&& $\!\!\!\!\!\!t^{\prime} \Delta t^2\!\!\!\!\!\!$ 
&& $\!\!\!\!\!\!0\!\!\!\!\!\!$ && $0$ & \cr
\omit&height2pt&\omit&&\omit&&\omit&&\omit&\cr \tablerule
\omit&height2pt&\omit&&\omit&&\omit&&\omit&\cr 
&& $\!\!\!\!\!\!5\!\!\!\!\!\!$ 
&& $\!\!\!\!\!\! -t^{\prime} \Delta t^2\!\!\!\!\!\!$ 
&& $\!\!\!\!\!\! -\frac23 t^{\prime 3} \Delta t +\frac32t^{\prime 2} 
\Delta t^2 \!\!\!\!\!\! $ && $-\frac1{3} t^{\prime 3} \Delta t^2$ & \cr
\omit&height2pt&\omit&&\omit&&\omit&&\omit&\cr \tablerule
\omit&height2pt&\omit&&\omit&&\omit&&\omit&\cr 
&& $\!\!\!\!\!\!6\!\!\!\!\!\!$ 
&& $\!\!\!\!\!\!0\!\!\!\!\!\!$ 
&& $\!\!\!\!\!\! -\frac13 t^{\prime 3} \Delta t +\frac12t^{\prime 2} 
\Delta t^2\!\!\!\!\!\!$ && $0$ & \cr
\omit&height2pt&\omit&&\omit&&\omit&&\omit&\cr \tablerule
\omit&height2pt&\omit&&\omit&&\omit&&\omit&\cr 
&& $\!\!\!\!\!\!7\!\!\!\!\!\!$ && $\!\!\!\!\!\!0\!\!\!\!\!\!$ 
&& $\!\!\!\!\!\!0\!\!\!\!\!\!$ && $0$ & \cr
\omit&height2pt&\omit&&\omit&&\omit&&\omit&\cr \tablerule
\omit&height2pt&\omit&&\omit&&\omit&&\omit&\cr 
&& $\!\!\!\!\!\! 8\!\!\!\!\!\!$ && $\!\!\!\!\!\!t^{\prime} \Delta t^2
\!\!\!\!\!\!$ && $\!\!\!\!\!\!\frac13 t^{\prime 3} \Delta t - t^{\prime 2}
\Delta t^2 \!\!\!\!\!\!$ && $0$ & \cr
\omit&height2pt&\omit&&\omit&&\omit&&\omit&\cr \tablerule
\omit&height2pt&\omit&&\omit&&\omit&&\omit&\cr 
&& $\!\!\!\!\!\! 9 \!\!\!\!\!\!$ && $\!\!\!\!\!\!0\!\!\!\!\!\!$ 
&& $\!\!\!\!\!\!\frac49 t^{\prime 3} \Delta t -\frac23 t^{\prime 2} 
\Delta t^2\!\!\!\!\!\!$ && $\frac2{9} t^{\prime 3} \Delta t^2$ & \cr
\omit&height2pt&\omit&&\omit&&\omit&&\omit&\cr \tablerule
\omit&height2pt&\omit&&\omit&&\omit&&\omit&\cr 
&& $\!\!\!\!\!\! 10 \!\!\!\!\!\!$ && $\!\!\!\!\!\!0\!\!\!\!\!\!$ 
&& $\!\!\!\!\!\!0\!\!\!\!\!\!$ && $0$ & \cr
\omit&height2pt&\omit&&\omit&&\omit&&\omit&\cr \tablerule
\omit&height1pt&\omit&&\omit&&\omit&&\omit&\cr \tablerule
\omit&height2pt&\omit&&\omit&&\omit&&\omit&\cr 
&& $\!\!\!\! {\rm Sum} \!\!\!\!\!\!\!$ && $\!\!\!\!\!\!t' \Delta t^2
\!\!\!\!\!\!$ && $\!\!\!\!\!\!\frac19 t^{\prime 3} \Delta t + \frac13 
t^{\prime 2} \Delta t^2 \!\!\!\!\!\!$ && $-\frac19 t^{\prime 3} 
\Delta t^2$ & \cr 
\omit&height2pt&\omit&&\omit&&\omit&&\omit&\cr \tablerule}}

\caption{Reductions involving $h_{0i}$. Each coefficient appears
in the form $\int_0^t \! dt' \! \int \! d\Omega \times {\rm Coef.} \times
f(t',\vec{x} \!+\! \Delta t \, \widehat{r})$.}

\label{h0i}

\end{table}

Although Tables~\ref{h00}-\ref{hijij} reduce $8\pi \delta \mathcal{P}(x)$
to a sum of surface terms roughly like those of $\delta \mathcal{V}(x)$,
we have still not reached an irreducible form from which a definitive
comparison can be made. The key to attaining such a form is to expand
the graviton fields in powers of $\Delta t$ and then perform the angular
integrations. The details of this procedure are explained in the Appendix
but the results for the three surface terms of expression (\ref{Volfin})
for $\delta \mathcal{V}(x)$ are simple enough to quote,
\begin{eqnarray}
\lefteqn{-\frac16 \! \int_0^t \!\! dt' (t^3 \!-\! \Delta t^3) \!
\int \!\! d\Omega \, h_{00}\Bigl(t',\vec{x} \!+\! \Delta t \, \widehat{r}
\Bigr) = \int_0^t \!\! dt' \Bigl[-\frac16 t^{\prime 3} \!-\!
\frac12 t^{\prime 2} \Delta t \!-\! \frac12 t' \Delta t^2\Bigr] } \nonumber \\
& & \hspace{2.7cm} \times \, 4\pi \!\! \sum_{n=0}^{\infty} \frac{\Delta t^{2n} 
\nabla^{2n}}{(2n \!+\! 1)!} \, h_{00}(t',\vec{x}) \; , \qquad \\
\lefteqn{\frac13 \! \int_0^t \!\! dt' (t^3 \!-\! \Delta t^3) \!
\int \!\! d\Omega \, \widehat{r}^i h_{0i}\Bigl(t',\vec{x} \!+\! \Delta t 
\, \widehat{r} \Bigr) = \int_0^t \!\! dt' \Bigl[\frac13 t^{\prime 3} \!+\!
t^{\prime 2} \Delta t \!+\! t' \Delta t^2\Bigr] } \nonumber \\
& & \hspace{2.7cm} \times \, 4\pi\!\! \sum_{n=0}^{\infty} \frac{\Delta t^{2n+1} 
\nabla^{2n}}{(2n \!+\! 1)! (2n \!+\! 3)} \, h_{0i , i}(t',\vec{x}) \; , 
\qquad \\
\lefteqn{-\frac16 \! \int_0^t \!\! dt' (t^3 \!-\! \Delta t^3) \!
\int \!\! d\Omega \, \widehat{r}^i \widehat{r}^j h_{ij}\Bigl(t',\vec{x} \!+\! 
\Delta t \, \widehat{r} \Bigr) = \int_0^t \!\! dt' \Bigl[-\frac16 
t^{\prime 3} \!-\! \frac12 t^{\prime 2} \Delta t \!-\! \frac12 t' 
\Delta t^2\Bigr] } \nonumber \\
& & \hspace{2.7cm} \times \, 4\pi \!\! \sum_{n=0}^{\infty} \frac{\Delta t^{2n} 
\nabla^{2n-2}}{(2n \!+\! 1)! (2n \!+\! 3)} \Bigl[ h_{ii,jj}(t',\vec{x}) 
\!+\! 2n h_{ij,ij}(t',\vec{x})\Bigr] \; . \qquad
\end{eqnarray}

Applying the same reduction to the terms of Tables~\ref{h00}-\ref{hijij},
and carrying out some judicious partial integrations with respect to $t'$,
allows us to reach a definitive expression for the difference of $8\pi 
\delta \mathcal{P}(x)$ and $\delta \mathcal{V}(x)$,
\begin{eqnarray}
\lefteqn{8\pi \delta \mathcal{P}(x) - \delta \mathcal{V}(x) = 
\int_0^t \!\! dt' \, t^{\prime 3} \Delta t^4 \times 4\pi \!\! 
\sum_{n=0}^{\infty} \frac{\Delta t^{2n} \nabla^{2n}}{(2n \!+\! 1)!
(2n \!+\! 3) (2n \!+\! 5)} } \nonumber \\
& & \hspace{.5cm} \times \Biggl\{\frac1{18} \nabla^4 h_{00}(t',\vec{x}) 
\!-\! \frac19 \nabla^2 \dot{h}_{0i,i}(t',\vec{x}) -\frac1{36} \nabla^2 
\ddot{h}_{ii}(t',\vec{x}) \nonumber \\
& & \hspace{2.8cm} + \frac1{36} \nabla^4 h_{ii}(t',\vec{x})
-\frac1{36} \nabla^2 h_{ij , ij}(t',\vec{x}) + \frac1{12}
\ddot{h}_{ij , ij}(t',\vec{x}) \Biggr\} . \qquad \label{defin}
\end{eqnarray}
The various graviton fields in (\ref{defin}) can be assembled into
components of the linearized curvature tensor,
\begin{eqnarray}
\lefteqn{\frac1{18} \nabla^4 h_{00} \!-\! \frac19 \nabla^2 \dot{h}_{0i,i}
\!-\! \frac1{36} \nabla^2 \ddot{h}_{ii} \!+\! \frac1{36} \nabla^4 h_{ii}
\!-\! \frac1{36} \nabla^2 h_{ij , ij} \!+\! \frac1{12} \ddot{h}_{ij , ij} } 
\nonumber \\
& & \hspace{0cm} = -\frac19 \nabla^2 \Bigl[ h_{0i , 0i} \!-\! \frac12 
h_{00 , ii} \!-\! \frac12 h_{ii , 00}\Bigr] \nonumber \\
& & \hspace{4.3cm} -\frac1{36} \nabla^2 \Bigl[
h_{ij , ij} \!-\! h_{ii , jj}\Bigr] + \frac1{12} \partial_0^2
\Bigl[ h_{ij , ij} \!-\! h_{ii , jj}\Bigr] \; , \qquad \\
& & \hspace{0cm} = \frac1{18} \nabla^2 \delta R - \frac1{12} \partial^2
\delta R_{ij ij} \; . 
\end{eqnarray}
Hence our final result takes the form,
\begin{eqnarray}
\lefteqn{8\pi \delta \mathcal{P}(x) - \delta \mathcal{V}(x) = 
\int_0^t \!\! dt' \, t^{\prime 3} \Delta t^4 \times 4\pi \!\! 
\sum_{n=0}^{\infty} \frac{\Delta t^{2n} \nabla^{2n}}{(2n \!+\! 1)!
(2n \!+\! 3) (2n \!+\! 5)} } \nonumber \\
& & \hspace{3.5cm} \times \Biggl\{ \frac1{18} \nabla^2 \delta R(t',\vec{x})
-\frac1{12} \Bigl[-\partial_0^{\prime 2} \!+\! \nabla^2\Bigr] \delta
R_{ijij}(t',\vec{x}) \Biggr\} . \qquad \label{finalcomp}
\end{eqnarray}

\begin{table}

\vbox{\tabskip=0pt \offinterlineskip
\def\tablerule{\noalign{\hrule}}
\halign to390pt {\strut#& \vrule#\tabskip=1em plus2em&
\hfil#\hfil& \vrule#& \hfil#\hfil& \vrule#& \hfil#\hfil&
\vrule#\tabskip=0pt\cr \tablerule
\omit&height4pt&\omit&&\omit&&\omit&\cr
\omit&height2pt&\omit&&\omit&&\omit&\cr 
&&\omit\hidewidth $\;{\rm \#}$ \hidewidth 
&& $\!\!\!\!{\rm Coef.\ of} \; h_{ii} \!\!\!\!$ 
&& $\!\!\!\!{\rm Coef.\ of} \; \widehat{r}^j h_{ii,j} \!\!\!\!$ & \cr
\omit&height4pt&\omit&&\omit&&\omit&\cr \tablerule 
\omit&height2pt&\omit&&\omit&&\omit&\cr 
&& 1 && $-\frac16 t^{\prime 3} + \frac12 t^{\prime 2} \Delta t$ 
&& $-\frac16 t^{\prime 3} \Delta t$ & \cr
\omit&height2pt&\omit&&\omit&&\omit&\cr \tablerule
\omit&height2pt&\omit&&\omit&&\omit&\cr 
&& 2 && $0$ && $0$ & \cr
\omit&height2pt&\omit&&\omit&&\omit&\cr \tablerule
\omit&height2pt&\omit&&\omit&&\omit&\cr 
&& 3 && $-\frac12 t^{\prime } \Delta t^2$ && $0$ & \cr
\omit&height2pt&\omit&&\omit&&\omit&\cr \tablerule
\omit&height2pt&\omit&&\omit&&\omit&\cr 
&& 4 && $0$ && $0$ & \cr
\omit&height2pt&\omit&&\omit&&\omit&\cr \tablerule
\omit&height2pt&\omit&&\omit&&\omit&\cr 
&& 5 && $0$ && $0$ & \cr
\omit&height2pt&\omit&&\omit&&\omit&\cr \tablerule
\omit&height2pt&\omit&&\omit&&\omit&\cr 
&& 6 && $0$ && $0$ & \cr
\omit&height2pt&\omit&&\omit&&\omit&\cr \tablerule
\omit&height2pt&\omit&&\omit&&\omit&\cr 
&& 7 && $\frac13 t^{\prime 3} - 2 t^{\prime 2} \Delta t + t^{\prime } \Delta t^2$ 
&& $\frac13 t^{\prime 3} \Delta t -\frac12 t^{\prime 2} \Delta t^2$ & \cr
\omit&height2pt&\omit&&\omit&&\omit&\cr \tablerule
\omit&height2pt&\omit&&\omit&&\omit&\cr 
&& 8 && $0$ && $0$ & \cr
\omit&height2pt&\omit&&\omit&&\omit&\cr \tablerule
\omit&height2pt&\omit&&\omit&&\omit&\cr 
&& 9 && $0$ && $0$ & \cr
\omit&height2pt&\omit&&\omit&&\omit&\cr \tablerule
\omit&height2pt&\omit&&\omit&&\omit&\cr 
&& 10 && $-\frac29 t^{\prime 3} + \frac43 t^{\prime 2} \Delta t -\frac23 t^{\prime } \Delta t^2$ 
&& $-\frac29 t^{\prime 3} \Delta t +\frac23 t^{\prime 2} \Delta t^2$ & \cr
\omit&height2pt&\omit&&\omit&&\omit&\cr \tablerule
\omit&height1pt&\omit&&\omit&&\omit&\cr \tablerule
\omit&height2pt&\omit&&\omit&&\omit&\cr 
&& Sum && $-\frac1{18} t^{\prime 3} -\frac16 t^{\prime 2} \Delta t -\frac16
t' \Delta t^2$ && $-\frac1{18} t^{\prime 3} \Delta t + \frac16 t^{\prime 2}
\Delta t^2$ & \cr 
\omit&height2pt&\omit&&\omit&&\omit&\cr \tablerule}}

\caption{Reductions involving $h_{ii}$. Each coefficient appears
in the form $\int_0^t \! dt' \! \int \! d\Omega \times {\rm Coef.} \times
f(t',\vec{x} \!+\! \Delta t \, \widehat{r})$.}

\label{hii}

\end{table}

\begin{table}

\vbox{\tabskip=0pt \offinterlineskip
\def\tablerule{\noalign{\hrule}}
\halign to390pt {\strut#& \vrule#\tabskip=1em plus2em&
\hfil#\hfil& \vrule#& \hfil#\hfil& \vrule#& \hfil#\hfil&
\vrule#\tabskip=0pt\cr \tablerule
\omit&height4pt&\omit&&\omit&&\omit&\cr
\omit&height2pt&\omit&&\omit&&\omit&\cr 
&&\omit\hidewidth $\;{\rm \#}$ \hidewidth 
&& $\!\!\!\!{\rm Coef.\ of} \; \widehat{r}^i h_{ij,j} \!\!\!\!$ 
&& $\!\!\!\!{\rm Coef.\ of} \; h_{ij,ij} \!\!\!\!$ & \cr
\omit&height4pt&\omit&&\omit&&\omit&\cr \tablerule 
\omit&height2pt&\omit&&\omit&&\omit&\cr 
&& 1 && $0$ && $0$ & \cr
\omit&height2pt&\omit&&\omit&&\omit&\cr \tablerule
\omit&height2pt&\omit&&\omit&&\omit&\cr 
&& 2 && $0$ && $0$ & \cr
\omit&height2pt&\omit&&\omit&&\omit&\cr \tablerule
\omit&height2pt&\omit&&\omit&&\omit&\cr 
&& 3 && $0$ && $0$ & \cr
\omit&height2pt&\omit&&\omit&&\omit&\cr \tablerule
\omit&height2pt&\omit&&\omit&&\omit&\cr 
&& 4 && $0$ && $0$ & \cr
\omit&height2pt&\omit&&\omit&&\omit&\cr \tablerule
\omit&height2pt&\omit&&\omit&&\omit&\cr 
&& 5 && $-\frac12 t^{\prime 2} \Delta t^2$ 
&& $\frac16 t^{\prime 3} \Delta t^2$ & \cr
\omit&height2pt&\omit&&\omit&&\omit&\cr \tablerule
\omit&height2pt&\omit&&\omit&&\omit&\cr 
&& 6 && $0$ && $0$ & \cr
\omit&height2pt&\omit&&\omit&&\omit&\cr \tablerule
\omit&height2pt&\omit&&\omit&&\omit&\cr 
&& 7 && $0$ && $0$ & \cr
\omit&height2pt&\omit&&\omit&&\omit&\cr \tablerule
\omit&height2pt&\omit&&\omit&&\omit&\cr 
&& 8 && $0$ && $0$ & \cr
\omit&height2pt&\omit&&\omit&&\omit&\cr \tablerule
\omit&height2pt&\omit&&\omit&&\omit&\cr 
&& 9 && $0$ 
&& $-\frac19 t^{\prime 3} \Delta t^2$ & \cr
\omit&height2pt&\omit&&\omit&&\omit&\cr \tablerule
\omit&height2pt&\omit&&\omit&&\omit&\cr 
&& 10 && $0$ && $0$ & \cr
\omit&height2pt&\omit&&\omit&&\omit&\cr \tablerule
\omit&height1pt&\omit&&\omit&&\omit&\cr \tablerule
\omit&height2pt&\omit&&\omit&&\omit&\cr 
&& Sum && $-\frac12 t^{\prime 2} \Delta t^2$ 
&& $\frac1{18} t^{\prime 3} \Delta t^2$ &\cr
\omit&height2pt&\omit&&\omit&&\omit&\cr \tablerule}}

\caption{Reductions involving $h_{ij,j}$. Each coefficient appears
in the form $\int_0^t \! dt' \! \int \! d\Omega \times {\rm Coef.} \times
f(t',\vec{x} \!+\! \Delta t \, \widehat{r})$.}

\label{hijij}

\end{table}

\section{Discussion}

The invariant volume of the past light-cone is an interesting
quantity because it provides a partial solution to the tough
problem of constructing observables for quantum gravity \cite{TW1,obs}, 
because it can play a role in characterizing the quantum field theoretic
back-reaction on spacetime expansion \cite{obs,TW2}, and because its 
gradient can provide an alternative to the timelike vector field 
involved in certain alternate gravity models \cite{Ted,JB} without 
introducing new dynamical degrees of freedom. It is well known that 
nonlocal functionals of the metric arise from quantum corrections to 
the effective field equations and a number of authors have considered 
nonlocal gravity models \cite{TW3,DW,nonloc}.

We have studied the relation between the invariant volume of the 
past light-cone $\mathcal{V}[g](x)$ and the Paneitz operator $D_P$,
a 4th order differential operator which occurs in the theory of
conformal anomalies. Based on their flat space limits we conjectured
that acting $D_P$ on $\mathcal{V}[g](x)$ might give $8\pi$ for a
general metric. We checked this conjecture in its integral form by
comparing $\mathcal{V}[g](x)$ with $8\pi$ times $\mathcal{P}[g](x)$,
the integral of the retarded Green's function of the Paneitz operator.
If the same operator whose logarithm occurs in the ubiquitous conformal
anomalies \cite{SD1,SD2} could be shown to give the invariant volume
of the past light-cone then alternate gravity models which involve the 
latter would become considerably more plausible.

Section 2 considered the case of an arbitrary homogeneous and
iso\-tro\-pic geometry, which has great significance for cosmology. 
We explicitly constructed the invariant volume of the past light-cone 
(\ref{Vbar}) and $8\pi$ times the integral of the Paneitz Greens function 
(\ref{Pbar}). Some trivial calculus manipulations suffice to show that 
the two expressions agree exactly for the case of zero or negative
spatial curvature. For positive spatial curvature the two expressions
agree when the observation point occurs less that one Hubble time 
later than the initial value surface. After one Hubble time $\mathcal{V}$
does not agree with $8\pi \mathcal{P}$ unless one modifies $\mathcal{V}$
to be the sum of the volumes of the past light-cone from the observation
point and from every focal point at which past-directed, null geodesics
from the observation point converge.

In section 3 we compared $\mathcal{V}[\eta+h](x)$ with $8\pi 
\mathcal{P}[\eta+h](x)$ at first order in perturbation theory about flat 
spacetime. An explicit expression (\ref{Volfin}) was derived for 
$\delta \mathcal{V}(x)$, and another expression (\ref{Panfin}) was 
obtained for $8\pi \delta \mathcal{P}(x)$. It was not so easy to compare 
the two relations but we eventually obtained a definitive result 
(\ref{finalcomp}) for their difference. Although expression 
(\ref{finalcomp}) is not zero, it does vanish for an arbitrary 
linearized solution of the vacuum Einstein equations because they imply,
\begin{equation}
R_{\mu\nu} - \frac12 g_{\mu\nu} R = 0 \qquad \Longrightarrow \qquad
\delta R = 0 \quad {\rm and} \quad \partial^2 \delta R_{\rho\sigma\mu\nu} 
= 0 \; .
\end{equation}

We do not yet know what the vanishing of (\ref{finalcomp}) with the
linearized Einstein equations means. That $8\pi \delta \mathcal{P}(x) -
\delta \mathcal{V}(x)$ must involve the linearized curvature tensor
follows because $\mathcal{V}[\eta+h](x)$ and $8\pi \mathcal{P}[\eta+h](x)$
agree for $h_{\mu\nu} = 0$, and both transform as scalars under any 
diffeomorphism which preserves the initial value surface $\mathcal{S}$. 
However, not all components of the linearized curvature tensor vanish 
with the linearized Einstein equations --- for example, $\delta R_{ijij}$ 
does not, nor does $\delta R_{0i0i}$. Yet only vanishing combinations
appeared in the difference (\ref{finalcomp}). This seems unlikely to
have been an accident, but we do not understand its significance.

One might wonder if $D_P$ can be changed by some local operator 
to make the difference (\ref{finalcomp}) go away. The answer is no.
If there were such an operator then acting $\partial^4$ on 
(\ref{finalcomp}) would give this operator acting on $t^4/24$.
However, direct computation shows that acting $\partial^4$ on a
nonlocal expression of the form (\ref{finalcomp}) fails to localize
it,
\begin{eqnarray}
\lefteqn{\partial^4 \int_0^t \!\! dt' \, t^{\prime 3} \Delta t^4 
\times 4\pi \!\! \sum_{n=0}^{\infty} \frac{\Delta t^{2n} 
\nabla^{2n}}{(2n \!+\! 1)! (2n \!+\! 3) (2n \!+\! 5)} \, f(t',\vec{x}) } 
\nonumber \\
& & \hspace{2.5cm} = \int_0^t \!\! dt' \, t^{\prime 3} 
\times 4\pi \!\! \sum_{n=0}^{\infty} \frac{\Delta t^{2n} 
\nabla^{2n}}{(2n \!+\! 1)! (2n \!+\! 3) (2n \!+\! 5)} \, f(t',\vec{x}) 
\; . \qquad
\end{eqnarray}

\vskip .5cm

\centerline{\bf Acknowledgements}

It is a pleasure to acknowledge conversations and correspondence
on this subject with C. Deffayet, S. Deser, G. Esposito-Farese
and N. C. Tsamis. This work was partially supported by NSF grants 
PHY-0653085 and PHY-0855021 and by the Institute for Fundamental 
Theory at the University of Florida.

\section{Appendix}

The purpose of this appendix is to derive some relations which apply to
the angular integral of functions over the surface of the flat space 
light-cone. One can represent such a function as $f(\vec{x} \!+\! \Delta t \, 
\widehat{r})$, and the relations all derive from expanding in powers of
$\Delta t$,
\begin{equation}
f(\vec{x} \!+\! \Delta t \, \widehat{r}) = \sum_{n=0}^{\infty} 
\frac{\Delta t^n}{n!} \, (\widehat{r} \!\cdot\! \vec{\nabla})^n f(\vec{x})
\; . 
\end{equation}
This brings all factors of the unit vector $\widehat{r}$ outside the
function, whereupon we can evaluate the angular integrations using 
the relation,
\begin{equation}
\int \!\! d\Omega \, \widehat{r}^{i_1} \widehat{r}^{i_2} \cdots
\widehat{r}^{i_n} = 4\pi \cases{ 0 & n \, {\rm odd} \cr
\frac1{n\!+\!1} \, \delta^{(i_1 i_2} \cdots 
\delta^{i_{n-1} i_n)} & n \, {\rm even} \cr} \; .
\end{equation}
The reductions of section 3.3 necessitate consideration of
$f(\vec{x} \!+\! \Delta t \, \widehat{r})$ by itself, or multiplied
with up to three unit vectors,
\begin{eqnarray}
\int \!\! d\Omega \, f(\vec{x} \!+\! \Delta t \, \widehat{r}) &\!\!\!\!\!
= \!\!\!\!\! & 4\pi \!\! \sum_{n=0}^{\infty} \frac{ \Delta t^{2n} \nabla^{2n}}{
(2n \!+\! 1)!} \, f(\vec{x}) \; , \\
\int \!\! d\Omega \, \widehat{r}^i f(\vec{x} \!+\! \Delta t \, \widehat{r}) 
& \!\!\!\!\! = \!\!\!\!\! & 4\pi \!\!\sum_{n=0}^{\infty} \frac{ \Delta t^{2n+1} 
\nabla^{2n}}{(2n \!+\! 1)! (2n \!+\!3)} \, \partial_i f(\vec{x}) \; , \\
\int \!\! d\Omega \, \widehat{r}^i \widehat{r}^j f(\vec{x} \!+\! \Delta t 
\, \widehat{r}) & \!\!\!\!\! = \!\!\!\!\! & 4\pi \!\! \sum_{n=0}^{\infty} 
\frac{ \Delta t^{2n} [\delta^{ij} \nabla^{2n} \!\!+\! 2n \partial^i \partial^j 
\nabla^{2n-2}]}{(2n \!+\! 1)! (2n \!+\!3)} \, f(\vec{x}) \; , \\
\int \!\! d\Omega \, \widehat{r}^i \widehat{r}^j \widehat{r}^k f(\vec{x} 
\!+\! \Delta t \, \widehat{r}) & \!\!\!\!\! = \!\!\!\!\! & 4\pi \!\!
\sum_{n=0}^{\infty} \frac{\Delta t^{2n+1} [3 \delta^{(ij} \partial^{k)} 
\nabla^{2n} \!\!+\! 2n \partial^i \partial^j \partial^k \nabla^{2n-2}]}{(2n 
\!+\! 1)! (2n \!+\!3) (2n \!+\! 5)} \, f(\vec{x}) \; . \qquad
\end{eqnarray}
By combining and comparing these expressions one can derive the following
identities which were used in preparing Tables~\ref{h00}-\ref{hijij},
\begin{eqnarray}
\int \!\! d\Omega \, \Bigl[\nabla^2 - (\widehat{r} \!\cdot\!\! \vec{\nabla})^2
\Bigr] f(\vec{x} \!+\! \Delta t \, \widehat{r}) & = & \frac{2}{\Delta t}
\int \!\! d\Omega \, \widehat{r} \!\cdot\!\! \vec{\nabla}
f(\vec{x} \!+\! \Delta t \, \widehat{r}) \; , \\
\int \!\! d\Omega \, \Bigl[\nabla^2 - (\widehat{r} \!\cdot\!\! \vec{\nabla})^2
\Bigr] \widehat{r}^i f(\vec{x} \!+\! \Delta t \, \widehat{r}) & = & 
\frac{2}{\Delta t} \int \!\! d\Omega \, \Bigl[\partial_i - 
\frac{3 \widehat{r}^i}{\Delta t}\Bigr] f(\vec{x} \!+\! \Delta t \, 
\widehat{r}) \; , \qquad \\
\int \!\! d\Omega \, \Bigl[\partial_i - \widehat{r}^i \widehat{r} \!\cdot\!\!
\vec{\nabla} \Bigr] f_i(\vec{x} \!+\! \Delta t \, \widehat{r}) & = & 
\frac{2}{\Delta t} \int \!\! d\Omega \, \widehat{r}^i f_i(\vec{x} \!+\! 
\Delta t \, \widehat{r}) \; .
\end{eqnarray}

\end{document}